\documentclass[aps,prX,showpacs,twocolumn,longbibliography,amsmath,amssymb]{revtex4-1}
\usepackage{graphicx}
\usepackage{dcolumn}
\usepackage{bm}
\usepackage{amsmath}
\usepackage{amsfonts}
\usepackage{amssymb}
\usepackage{multirow}

\usepackage{amsthm}
\usepackage{pinlabel}

\usepackage{epstopdf}
\usepackage[abs]{overpic}

\usepackage{cancel}

\usepackage{lipsum}  

\usepackage[normalem]{ulem}

\usepackage[dvipsnames]{xcolor}

\usepackage{xr}

\usepackage{braket}

\definecolor{specialgray}{HTML}{505050}
\definecolor{col10K}{HTML}{FFA000}
\definecolor{col300K}{HTML}{924FA4}
\definecolor{colMu}{HTML}{5278BD}
\definecolor{colMuI}{HTML}{924FA4}
\definecolor{newred}{HTML}{D53E4F}
\definecolor{newblue}{HTML}{5278BD}
\definecolor{newcyan}{HTML}{4EBCB3}
\definecolor{newgreen}{HTML}{5CB14E}
\definecolor{newpurple}{HTML}{924FA4}
\definecolor{newyellow}{HTML}{D1C72E}
\definecolor{neworange}{HTML}{D6923C}

\usepackage{lineno}
\usepackage{placeins}

\begin{document}
\title{Induced odd-frequency superconducting state in vertex-corrected Eliashberg theory}
\author{Fabian Schrodi}\email{fabian.schrodi@physics.uu.se}
\author{Alex Aperis}\email{alex.aperis@physics.uu.se}
\author{Peter M. Oppeneer}\email{peter.oppeneer@physics.uu.se}
\affiliation{Department of Physics and Astronomy, Uppsala University, P.\ O.\ Box 516, SE-75120 Uppsala, Sweden}
	
\vskip 0.4cm
\date{\today}

\begin{abstract}
	\noindent 
	We show that vertex corrections to Migdal's theorem in general induce an odd-frequency spin-triplet superconducting order parameter, which coexists with its more commonly known even-frequency spin-singlet counterpart. Fully self-consistent vertex-corrected Eliashberg theory calculations for a two dimensional cuprate model, isotropically coupled to an Einstein phonon, confirm that both superconducting gaps are finite over a wide range of temperatures. The subordinate $d$-wave odd-frequency superconducting gap is found to be one order of magnitude smaller than the primary even-frequency $d$-wave gap. Our study provides a direct proof of concept for a previously unknown generation mechanism of odd-frequency superconductivity as well as for the generic	coexistence of both superconducting states in bulk materials.
\end{abstract}

\maketitle

\section{Introduction}\label{scIntro}

Odd-frequency superconductivity describes a Cooper pair condensate in which the quantum mechanical wave function is odd under exchange of relative time between two electrons of a Cooper pair. As first proposed by Berezinskii \cite{Berezinskii1974}, this ordered state is compatible with the Pauli principle as long as either the momentum space parity is even and the electrons are of spin-triplet type, or the parity is odd and one considers spin-singlet electron pairs, see e.g.\ Ref.\,\cite{Linder2019} for a recent review. This type of superconductivity is clearly beyond the Bardeen-Cooper-Schrieffer (BCS) theory that neglects time-retardation effects between the electrons of the Cooper pair, but can be described naturally within Eliashberg's theory \cite{Eliashberg1960}.
Historically, it has been argued that electron-phonon coupling can theoretically lead to  odd-frequency superconductivity, provided that the interaction has a sufficiently strong odd momentum component as a result of coupling between electrons and acoustic phonons \cite{Balatsky1992}. However, such a scenario was considered unlikely if no spin dependence enters into the interaction \cite{Abrahams1993}. More recent theories have shown that signatures of an odd-frequency phase are indeed possible due to electron-phonon interactions only, depending on the interaction strength \cite{Kusunose2011,Matsumoto2012,Aperis2015,Aperis2020}.

Odd-frequency superconductivity has been discussed recurrently in various forms \cite{Bergeret2005,Linder2019}. In this discussion it is important to distinguish between existence of a superconducting condensate with an odd-frequency order parameter, i.e., with nonzero gap function,  and odd-frequency pair correlations. The latter can occur when a frequency-even superconductor is placed in close proximity to a magnetic layer or when a magnetic impurity is placed on a conventional superconductor \cite{Bergeret2005,Yokoyama2007,Linder2008,Linder2010,Tanaka2012,Alidoust2014,Linder2015,Pal2017,Kuzmanovski2020}. The spins of the conventional Cooper-pair electrons become rotated in the exchange field, leading to a spin-triplet component with time-odd parity \cite{Bergeret2005,Eschrig2008}. Recent experiments have provided spectroscopic observations of odd-frequency pair correlations \cite{DiBernardo2015,Dibernardo2015b,Pal2017,Diesch2018,Perrin2020}. The observation of an odd-frequency superconductivity phase in a bulk material remains however thus far elusive. 
The thermodynamic stability of an odd-frequency phase in the presence 
of an even-frequency order parameter has therefore been studied theoretically in several works \cite{Belitz1999,Solenov2009,Kusunose2011b,Aperis2015,Aperis2020}.

In many of the theoretical proposals for systems exhibiting this rather exotic type of Cooper pairing, there exists one particular degree of freedom that makes the occurrence of an odd-frequency superconducting  pairing possible. %even in the presence}
As laid out in Ref.\,\cite{Linder2019}, odd-frequency superconductivity in bulk materials is often associated with spin degrees of freedom, or with the strong-coupling limit of electron-phonon interactions \cite{Kusunose2011,Kusunose2011b}.
Other examples are pairing effects due to multiple band or electron orbitals \cite{Blackschaffer2013,Komendova2015}, influence of magnetic fields \cite{Matsumoto2012,Aperis2015}, magnetic impurities \cite{Kuzmanovski2020}, proximity effects \cite{Blackschaffer2013_2,Cayao2018}, or due to edge-state induced ferromagnetic fluctuations \cite{Matsubara2021}.

Here we show that this conception is generally not complete. By deriving self-consistent and vertex-corrected Eliashberg equations, we prove that a finite odd-frequency spin-triplet order parameter can theoretically always coexist along with the `standard' even-frequency spin-singlet superconducting gap. Vertex corrections to the electron-phonon problem generally become important if the ratio of phonon to electron energy scale is not small \cite{Migdal1958}, and it recently has been shown that they are a potential candidate for producing unconventional symmetries in several classes of superconductors \cite{Schrodi2020_2}. Our current results suggest that the phenomenology of vertex corrections is even richer, in that they can also be responsible for odd-frequency superconducting states. We benchmark our theory by solving the Eliashberg equations for a Holstein-like Hamiltonian with Einstein phonon and {\em isotropic} electron-phonon coupling, resembling the characteristics of a prototypical cuprate superconductor. The odd-frequency $d$-wave state is induced by its even-frequency counterpart, which is the primary order parameter, and is approximately one order of magnitude smaller in amplitude. Our results suggest that the coexistence of both condensates is possible due to a mixing of momentum space representations, triggered by the inclusion of vertex corrections into our Eliashberg framework.

In the following we assume for simplicity that the spatial parity is always even, hence we want to describe the most common Brillouin zone (BZ) symmetries of the superconducting gap function, such as $s$-wave or $d$-wave. We thereby leave the straightforward generalization to more exotic $p$-wave or $f$-wave states for future analysis. Consequently, we are left with spin and time (frequency) as the only two degrees of freedom, since we do not include multiple orbitals or bands in our formalism. Then the possible superconducting order parameters are either even in frequency and spin singlet, or odd in frequency and spin triplet.

The remainder of this paper is organized as follows: We start Section \ref{scVertex} by introducing the general mathematical framework of our theory, and present the vertex-corrected Eliashberg equations, as they were developed in Ref.\,\cite{Schrodi2020_2}, in Section \ref{scEven}. We show in Section \ref{scInduced} that the results to those self-consistent equations always induce an odd-frequency order parameter, which describes Cooper pairs of spin-triplet electrons. The self-consistent set of nonadiabatic Eliashberg equations, including both even- and odd-frequency order parameters, is given in Section \ref{scCoexist}. We test our theory by using a one-band tight-binding description, which describes a prototypical cuprate superconductor. Section \ref{scCuprate} contains an analysis of our numerical results with respect to momentum and temperature dependencies. Finally, we conclude with a short summary and discussion in Section \ref{scSummary}.

\section{Vertex-corrected Eliashberg theory}\label{scVertex}

Let us assume that the atomic vibrations in the superconductor can approximately be described by a single-branch optical phonon mode with frequency $\Omega$. The purely electronic degree of freedom is characterized by a single-band dispersion $\xi_{\mathbf{k}}$, with $\mathbf{k}$ a BZ wave vector. We consider a momentum independent electron-phonon scattering strength $g_0$, hence, the Hamiltonian of the system can be expressed as
\begin{align}
H =& \sum_{\mathbf{k}}\xi_{\mathbf{k}}\Psi^{\dagger}_{\mathbf{k}}\hat{\rho}_3\Psi_{\mathbf{k}} + \hbar\Omega\sum_{\mathbf{q}}\big(b^{\dagger}_{\mathbf{q}}b_{\mathbf{q}} + \frac{1}{2}\big) \nonumber\\
& + g_0 \sum_{\mathbf{k},\mathbf{q}} u_{\mathbf{q}} \Psi^{\dagger}_{\mathbf{k}-\mathbf{q}}\hat{\rho}_3\Psi_{\mathbf{k}} ~,
\end{align}
with phonon displacement $u_{\mathbf{q}}=b^{\dagger}_{\mathbf{q}}+b_{-\mathbf{q}}$, Pauli matrices $\hat{\rho}_i$ and Nambu spinor $\Psi_{\mathbf{k}}^{\dagger}=\big(c^{\dagger}_{\mathbf{k},\uparrow},c_{-\mathbf{k},\downarrow}\big)$. We use $b_{\mathbf{q}}^{\dagger}$ and $c_{\mathbf{k},\sigma}^{\dagger}$ as phonon and electron creation operators, with $\sigma\in\{\uparrow,\downarrow\}$ labeling the electron spin.

At temperature $T$ we  define fermion and boson Matsubara frequencies as $\omega_m=\pi T(2m+1)$ and $q_l=2\pi Tl$, respectively, with $m,\,l\in\mathbb{Z}$. For sake of brevity,  we adopt from here on the four-momenta notation $k=(\mathbf{k},i\omega_m)$ and $q=(\mathbf{q},iq_l)$. For the Einstein phonon spectrum considered here the phonon propagator can be written as a simple Lorentzian in frequency space, $D_q=2\Omega/(\Omega^2+q_q^2)$. The electron Green's function obeys the Dyson equation $\hat{G}_k = \hat{G}^0_k + \hat{G}^0_k \hat{\Sigma}_k \hat{G}_k$, which equivalently can be written as
\begin{align}
\hat{G}_k = \big[\hat{\rho}_0 - \hat{G}^0_k\hat{\Sigma}_k\big]^{-1} \hat{G}_k^0 ~. \label{G}
\end{align}
In Eq.\,(\ref{G}), $\hat{\Sigma}_k$ is the electron self-energy, and the bare electron Green's function is defined as $\hat{G}_k^0 = [i\omega_k \hat{\rho}_0 - \xi_{k} \hat{\rho}_3]^{-1}$. The electron self-energy including the two lowest order Feynman diagrams for electron-phonon scattering can be expressed as
\begin{align}
\hat{\Sigma}_k &= T \sum_{k_1} V_{k-k_1} \hat{\rho}_3 \hat{G}_{k_1} \hat{\rho}_3 \nonumber\\
&+ T^2 \sum_{k_1,k_2} V_{k-k_1}V_{k_1-k_2} \hat{\rho}_3 \hat{G}_{k_1}\hat{\rho}_3 \hat{G}_{k_2}\hat{\rho}_3 \hat{G}_{k_2-k_1+k} \hat{\rho}_3 , \label{sigma_nad}
\end{align}
where the electron-phonon interaction is defined via $V_{q}=g_0^2 D_{q}$\,\cite{Grimaldi1995,Botti2002,Cappelluti2003,Schrodi2020_2}. The first-order vertex corrections are given by the second term in Eq.\ (\ref{sigma_nad}).

In the following Section \ref{scEven} we derive self-consistent Eliashberg equations describing the interacting state of our model system, where we follow closely the notation of Refs.\,\cite{Schrodi2020_2,SchrodiDWaveProj}. We show in Section \ref{scInduced} that our theory allows for an induced subordinate odd-frequency spin-triplet superconducting state. A complete mathematical framework for the coexistence of even- and odd-frequency Cooper pairs is given in Section \ref{scCoexist}.

\subsection{Even-frequency, spin-singlet Cooper pairs} \label{scEven}

The vast majority of superconductors exhibits the formation of spin-singlet Cooper pairs, where all quantities transform even along the real and Matsubara frequency axis. Within our formalism, such a state can conveniently be described by the electron self-energy ansatz
\begin{align}
\hat{\Sigma}_k = i\omega_k (1-Z_k) \hat{\rho}_0 + \chi_k\hat{\rho}_3 + \phi_k\hat{\rho}_1  ,\label{sigma}
\end{align}
where the electron mass renormalization $Z_k$, chemical potential renormalization $\chi_k$ and superconducting order parameter $\phi_k$ are introduced. Plugging Eq.\,(\ref{sigma}) into Eq.\,(\ref{G}) gives
\begin{align}
\hat{G}_k = \frac{i\omega_kZ_k}{\Theta_k} \hat{\rho}_0 + \frac{\xi_k+\chi_k}{\Theta_k}\hat{\rho}_3 + \frac{\phi_k}{\Theta_k}\hat{\rho}_1 ,
\end{align}
with determinant
\begin{align}
\Theta_k = \big[i\omega_kZ_k\big]^2 - \big[\xi_k+\chi_k\big]^2 - \phi_k^2  ~.
\end{align}
As short-hand notation we introduce the symbols
\begin{align}
\gamma_k^{(Z)}=\frac{\omega_kZ_k}{\Theta_k}~,~\gamma_k^{(\chi)}=\frac{\xi_k+\chi_k}{\Theta_k} ~,~ \gamma_k^{(\phi)}=\frac{\phi_k}{\Theta_k}  ~, \label{defgamma}
\end{align}
so that the electron Green's function can be written as
\begin{align}
\hat{G}_k = i\gamma_k^{(Z)} \hat{\rho}_0 + \gamma_k^{(\chi)}\hat{\rho}_3 + \gamma_k^{(\phi)}\hat{\rho}_1 . \label{G2}
\end{align}
As has been shown in Ref.\,\cite{Schrodi2020_2}, Eqs.\,(\ref{sigma_nad}) and (\ref{G2}) can be used to derive a set of self-consistent vertex-corrected Eliashberg equations, reading
\begin{align}
Z_k &= 1 - \frac{T}{\omega_k} \sum_{k_1} V_{k-k_1} \Big( \gamma^{(Z)}_{k_1} \nonumber\\
&~~~~~~~~~~~~~ + T \sum_{k_2} V_{k_1-k_2} \vec{\gamma}^T_{k_2} P^{(Z)}_{k_1} \vec{\gamma}_{k_3} \Big) , \label{z} \\
\chi_k &= T \sum_{k_1} V_{k-k_1} \Big( \gamma^{(\chi)}_{k_1} + T \sum_{k_2} V_{k_1-k_2} \vec{\gamma}^T_{k_2} P^{(\chi)}_{k_1} \vec{\gamma}_{k_3} \Big) , \label{chi}\\
\phi_k &= -T \sum_{k_1} V_{k-k_1} \Big( \gamma^{(\phi)}_{k_1} + T \sum_{k_2} V_{k_1-k_2} \vec{\gamma}^T_{k_2} P^{(\phi)}_{k_1} \vec{\gamma}_{k_3} \Big)  . \label{phi}
\end{align}
For brevity, the above equations are defined in terms of the vector $\vec{\gamma}_k^T=\big(\gamma_k^{(Z)}, \gamma_k^{(\chi)}, \gamma_k^{(\phi)}\big)$, and we use $k_3=k_2-k_1+k$. Additionally, we define matrices
\renewcommand{\arraystretch}{1.5}
\begin{align}
\setlength\arraycolsep{1.5pt}
P^{(Z)}_k &= \begin{pmatrix}
-\gamma_k^{(Z)}&  \gamma_k^{(\chi)}& \gamma_k^{(\phi)}   \\
\gamma_k^{(\chi)}& \gamma_k^{(Z)}& 0 \\
-\gamma_k^{(\phi)}& 0& -\gamma_k^{(Z)} 
\end{pmatrix} , \\
\setlength\arraycolsep{1.5pt}
P^{(\chi)}_k &= \begin{pmatrix}
-\gamma_k^{(\chi)}&  -\gamma_k^{(Z)}& 0 \\
-\gamma_k^{(Z)}& \gamma_k^{(\chi)}& -\gamma_k^{(\phi)} \\
0& -\gamma_k^{(\phi)}& -\gamma_k^{(\chi)} 
\end{pmatrix}, \\
\setlength\arraycolsep{1.5pt}
P^{(\phi)}_k &= \begin{pmatrix}
-\gamma_k^{(\phi)}&  0& -\gamma_k^{(Z)} \\
0& \gamma_k^{(\phi)}& \gamma_k^{(\chi)} \\
\gamma_k^{(Z)}& \gamma_k^{(\chi)}& -\gamma_k^{(\phi)} 
\end{pmatrix} .
\end{align}
The vertex-corrected Eliashberg Eqs.\,(\ref{z})--(\ref{phi}) exhibit richer phenomenology and more degrees of freedom than the standard theory, which only takes into account the lowest order Feynman diagram. For example, it has been shown by the current authors that superconductivity can be partially suppressed by vertex corrections\,\cite{Schrodi2020_2}, and can eventually lead to unconventional BZ symmetries of the pairing function\,\cite{SchrodiDWaveProj,SchrodiThFeAsN}.

\subsection{Induced odd-frequency spin-triplet state}\label{scInduced}

We now want to take a closer look at the electron self-energy. The Pauli matrices $\hat{\rho}_i$, $i\in\{0,1,2,3\}$, form a basis of the $2\times2$ Nambu space employed here. Notice, however, that the ansatz for $\hat{\Sigma}_k$ in Eq.\,(\ref{sigma}) does not include $\hat{\rho}_2$. The neglect of this channel is justified due to the gauge freedom of Eliashberg theory, i.e. any function proportional to $\hat{\rho}_2$ describing spin-singlet, even-frequency Cooper pairs can be interpreted as a complex phase of the superconducting gap function, which will differ from $\phi_k$ only by a proportionality factor. We are allowed to set the prefactor of $\hat{\rho}_2$ identically zero, since no experimentally observable effects arise from this complex phase, except in Josephson tunneling \cite{Josephson1969} which is of no interest here. This is the well-established picture for standard Eliashberg theory\,\cite{mitrovic,Carbotte2003}, and remains valid even upon the inclusion of vertex corrections.

Considering the self-energy expression of Eq.\,(\ref{sigma_nad}), we can define $\hat{\Sigma}_k=\hat{\Sigma}_k^{(1)}+\hat{\Sigma}_k^{(2)}$, where $\hat{\Sigma}_k^{(1)}$ and $\hat{\Sigma}_k^{(2)}$ correspond to contributions from the first and second order Feynman diagrams, respectively. From our ansatz Eq.\,(\ref{sigma}) it follows that the electron Green's function does not include any term proportional to $\hat{\rho}_2$, therefore we obtain
\begin{align}
\frac{1}{2}\mathrm{Tr}\big[ \hat{\rho}_2 \hat{\Sigma}_k^{(1)} \big] = 0 
\end{align}
from Eq.\,(\ref{sigma_nad}), as expected. However, applying the same operation to the non-adiabatic correction yields
\begin{align}
\frac{1}{2}\mathrm{Tr}\big[ \hat{\rho}_2 \hat{\Sigma}_k^{(2)} \big] =& T^2 \sum_{k_1,k_2} V_{k-k_1} V_{k_1-k_2} \Big[\gamma_{k_1}^{(Z)}\big(\gamma_{k_2}^{(\phi)} \gamma_{k_3}^{(\chi)} \nonumber \\
+&\gamma_{k_2}^{(\chi)} \gamma_{k_3}^{(\phi)}\big)  + \gamma_{k_1}^{(\chi)}\big(\gamma_{k_2}^{(Z)} \gamma_{k_3}^{(\phi)} - \gamma_{k_2}^{(\phi)} \gamma_{k_3}^{(Z)}\big) \nonumber\\
-&\gamma_{k_1}^{(\phi)}\big(\gamma_{k_2}^{(Z)} \gamma_{k_3}^{(\chi)}
+ \gamma_{k_2}^{(\chi)} \gamma_{k_3}^{(Z)}\big) \Big]\equiv \zeta_k^{\mathrm{ind}} , \label{zetainduced}
\end{align}
which does generally not vanish. Due to the fact that $\zeta_k^{\mathrm{ind}}\propto\hat{\rho}_2$, even though our definition of $\hat{\Sigma}_k$ does not include the $\hat{\rho}_2$ channel, Eq.\,(\ref{zetainduced}) describes an induced state. Similar induction mechanisms due to magnetic fields have been previously proposed within purely BCS pictures \cite{Aperis2008,Aperis2010}, however here retardation effects, that are captured by Eliashberg theory, are crucial for discussing induced odd-frequency superconductivity \cite{Aperis2020}.

A closer inspection of the induced function $\zeta_k^{\mathrm{ind}}$ reveals that each term in Eq.\,(\ref{zetainduced}) contains $\gamma^{(Z)}$, $\gamma^{(\chi)}$ and $\gamma^{(\phi)}$ exactly once, with different combinations of momentum and frequency dependencies. Consequently, $\zeta_k^{\mathrm{ind}}=0$ in the normal state, because then $\phi_k$ and therefore $\gamma^{(\phi)}$ identically vanishes, compare Eq.\,(\ref{defgamma}). We can therefore safely conclude that $\zeta_k^{\mathrm{ind}}$ describes physics of the superconducting state.

Next, we want to examine the frequency symmetry of $\zeta_k^{\mathrm{ind}}$, which we do here by neglecting the momentum dependence for the moment. We can safely assume that $Z_m$ is even along the Matsubara frequency axis, therefore $Z_m=Z_{-m-1}$, and likewise for $\chi_m$ and $\phi_m$. From the definitions in Eq.\,(\ref{defgamma}) it directly follows that $\gamma^{(Z)}_m=-\gamma^{(Z)}_{-m-1}$, $\gamma^{(\chi)}_m=\gamma^{(\chi)}_{-m-1}$ and $\gamma^{(\phi)}_m=\gamma^{(\phi)}_{-m-1}$. For the sake of the argument, let us define
\begin{align}
\beta_m = \sum_{m_1,m_2}V_{m-m_1}V_{m_1-m_2} \gamma_{m_1}^{(Z)}\gamma_{m_2}^{(\chi)}\gamma_{m_2-m_1+m}^{(\phi)} , \label{beta1}
\end{align}
which is representative for each term in Eq.\,(\ref{zetainduced}). Note, that we write the definition of $m_3=m_2-m_1+m$ here explicitly. The function $\beta_m$ has fermionic Matsubara frequency symmetry, so we need to consider $\beta_{-m-1}$. Replacing $m$ with $-m-1$ in Eq.\,(\ref{beta1}) changes $V_{m-m_1}$ to $V_{-m_1-m-1}$ and $\gamma_{m_2-m_1+m}^{(\phi)}$ to $\gamma_{m_2-m_1-m-1}^{(\phi)}$. We therefore make the definition $\tilde{m}_1=-m_1-1$, or, equivalently, $m_1=-\tilde{m}_1-1$. Since the summation over index $m_1$ includes all integers, we alternatively sum over index $\tilde{m}_1$, so as to get
\begin{align}
\beta_{-m-1} = \sum_{\tilde{m}_1,m_2}V_{\tilde{m}_1-m}V_{-\tilde{m}_1-1-m_2}\nonumber\\
\times \gamma_{-\tilde{m}_1-1}^{(Z)}\gamma_{m_2}^{(\chi)}\gamma_{m_2+\tilde{m}_1-m}^{(\phi)} . \label{beta2}
\end{align}
Next, we make a similar definition as before, $\tilde{m}_2=-m_2-1$, leading to
\begin{align}
\beta_{-m-1} =& \sum_{\tilde{m}_1,\tilde{m}_2}V_{\tilde{m}_1-m}V_{\tilde{m}_2-\tilde{m}_1} \nonumber\\
&\times \gamma_{-\tilde{m}_1-1}^{(Z)}\gamma_{-\tilde{m}_2-1}^{(\chi)}\gamma_{\tilde{m}_1-\tilde{m}_2-m-1}^{(\phi)} . \label{beta3}
\end{align}
As a last step we invert the frequency index of each function on the second line of Eq.\,(\ref{beta3}), using the aforementioned symmetries. After renaming the dummy indices $\tilde{m}_1$ and $\tilde{m}_2$ to $m_1$ and $m_2$, respectively, we arrive at
\begin{align}
\beta_m=-\beta_{-m-1} ~.
\end{align}
From here it is trivial to show that all terms in Eq.\,(\ref{zetainduced}) obey this particular symmetry, hence $\zeta_k^{\mathrm{ind}}$ is an odd-frequency pairing function. Note, that this does not imply a breaking of time reversal symmetry \cite{Kuzmanovski2017,Linder2019}.

When focusing on the momentum dependence, it is more difficult to make concrete analytical statements. This is due to the fact that the BZ symmetries of $\phi$ and $\chi$ are \textit{a priori} not known. We return to this aspect in Section \ref{scMom} when presenting our numerical results. 

As mentioned before, we assume even momentum space parity and neglect orbital/band degrees of freedom in this work, hence we can restrict our symmetry classification of $\zeta_k^{\mathrm{ind}}$ to spin and frequencies. We have shown analytically that the induced function is odd along the Matsubara frequency axis, it therefore describes spin triplet electron pairs\,\cite{Linder2019} which coexist with the even-frequency, spin singlet Cooper pairs described by $\phi_k$. Needless to say, all this interpretation is needed only if $\zeta_k^{\mathrm{ind}}$ is finite, which can not be guaranteed from the functional form of Eq.\,(\ref{zetainduced}). Therefore, we derive in the following Section \ref{scCoexist} the self-consistent equations governing the coexistence of superconducting states $\phi_k$ and $\zeta_k$. This is a necessary step for showing that the vertex-corrected electron-phonon interaction can support the spin-triplet odd-frequency state.

\subsection{Equations for coexisting pairing channels}\label{scCoexist}

We repeat the derivation of Section \ref{scEven} in the most general way possible within the $2\times2$ Nambu space employed here. The electron self-energy given by Eq.\,(\ref{sigma_nad}) remains valid, while we modify the ansatz in Eq.\,(\ref{sigma}) to
\begin{align}
\hat{\Sigma}_k= i\omega_k (1-Z_k)\hat{\rho}_0 + \chi_k\hat{\rho}_3 + \phi_k\hat{\rho}_1 + \zeta_k\hat{\rho}_2 , \label{sigma_general}
\end{align}
i.e., we now explicitly include the $\hat{\rho}_2$ channel. At this stage $\zeta_k$ is just a mathematical definition, and we analyze its meaning later. With Eq.\,(\ref{sigma_general}) at hand, the electron Green's function reads
\begin{align}
\hat{G}_k = i\gamma_k^{(Z)}\hat{\rho}_0 + \gamma_k^{(\chi)}\hat{\rho}_3 + \gamma_k^{(\phi)}\hat{\rho}_1 + \gamma_k^{(\zeta)}\hat{\rho}_{2} ,
\end{align}
where we use the straight-forward definitions $\gamma_k^{(\zeta)}=\zeta_k/\Theta_k$ and 
\begin{align}
\Theta_k = \big[i\omega_kZ_k\big]^2 - \big[\xi_k+\chi_k\big]^2 - \phi_k^2 - \zeta_k^2 .
\end{align}

Due to the self-consistent inclusion of $\zeta_k$ in our Eliashberg formalism, we get a total of four equations, reading
\begin{align}
Z_k &= 1 - \frac{T}{\omega_k} \sum_{k_1} V_{k-k_1} \Big( \gamma^{(Z)}_{k_1} \nonumber\\
&~~~~~~~~~~~~~ + T \sum_{k_2} V_{k_1-k_2} \vec{\gamma}^T_{k_2} Q^{(Z)}_{k_1} \vec{\gamma}_{k_3} \Big), \label{fullzz}\\
\chi_k &= T \sum_{k_1} V_{k-k_1} \Big( \gamma^{(\chi)}_{k_1} + T \sum_{k_2} V_{k_1-k_2} \vec{\gamma}^T_{k_2} Q^{(\chi)}_{k_1} \vec{\gamma}_{k_3} \Big) , \label{fullchi}\\
\phi_k &= -T \sum_{k_1} V_{k-k_1} \Big( \gamma^{(\phi)}_{k_1} + T \sum_{k_2} V_{k_1-k_2} \vec{\gamma}^T_{k_2} Q^{(\phi)}_{k_1} \vec{\gamma}_{k_3} \Big), \label{fullphi} \\
\zeta_k &= -T \sum_{k_1} V_{k-k_1} \Big( \gamma^{(\zeta)}_{k_1} + T \sum_{k_2} V_{k_1-k_2} \vec{\gamma}^T_{k_2} Q^{(\zeta)}_{k_1} \vec{\gamma}_{k_3} \Big) . \label{fullzeta}
\end{align}
Instead of the three-component vectors used in Section \ref{scEven}, Eqs.\,(\ref{fullzz})--(\ref{fullzeta}) are defined in terms of  $\vec{\gamma}_k^T=\big(\gamma_k^{(Z)}, \gamma_k^{(\chi)}, \gamma_k^{(\phi)} ,\gamma_k^{(\zeta)} \big)$. Likewise, the matrices $Q_k^{(\cdot)}$ are extended to a $4\times4$ pseudovector space, and read
\renewcommand{\arraystretch}{1.5}
\begin{align}
\setlength\arraycolsep{1.5pt}
Q^{(Z)}_k = \begin{pmatrix}
-\gamma_k^{(Z)}&  \gamma_k^{(\chi)}& \gamma_k^{(\phi)} & \gamma_k^{(\zeta)}  \\
\gamma_k^{(\chi)}& \gamma_k^{(Z)}& \gamma_k^{(\zeta)} & -\gamma_k^{(\phi)} \\
-\gamma_k^{(\phi)}& \gamma_k^{(\zeta)}& -\gamma_k^{(Z)} & -\gamma_k^{(\chi)} \\
-\gamma_k^{(\zeta)}& -\gamma_k^{(\phi)}& \gamma_k^{(\chi)} & -\gamma_k^{(Z)}
\end{pmatrix} ,
\end{align}
\begin{align}
\setlength\arraycolsep{1.5pt}
Q^{(\chi)}_k = \begin{pmatrix}
-\gamma_k^{(\chi)}&  -\gamma_k^{(Z)}& \gamma_k^{(\zeta)} & -\gamma_k^{(\phi)}  \\
-\gamma_k^{(Z)}& \gamma_k^{(\chi)}& -\gamma_k^{(\phi)} & -\gamma_k^{(\zeta)} \\
-\gamma_k^{(\zeta)}& -\gamma_k^{(\phi)}& -\gamma_k^{(\chi)} & \gamma_k^{(Z)} \\
\gamma_k^{(\phi)}& -\gamma_k^{(\zeta)}& -\gamma_k^{(Z)} & -\gamma_k^{(\chi)}
\end{pmatrix}, \\
\setlength\arraycolsep{1.5pt}
Q^{(\phi)}_k = \begin{pmatrix}
-\gamma_k^{(\phi)}&  -\gamma_k^{(\zeta)}& -\gamma_k^{(Z)} & \gamma_k^{(\chi)}  \\
-\gamma_k^{(\zeta)}& \gamma_k^{(\phi)}& \gamma_k^{(\chi)} & \gamma_k^{(Z)} \\
\gamma_k^{(Z)}& \gamma_k^{(\chi)}& -\gamma_k^{(\phi)} & \gamma_k^{(\zeta)} \\
-\gamma_k^{(\chi)}& \gamma_k^{(Z)}& -\gamma_k^{(\zeta)} & -\gamma_k^{(\phi)}
\end{pmatrix} ,\\
\setlength\arraycolsep{1.5pt}
Q^{(\zeta)}_k = \begin{pmatrix}
-\gamma_k^{(\zeta)}&  \gamma_k^{(\phi)}& -\gamma_k^{(\chi)} & -\gamma_k^{(Z)}  \\
\gamma_k^{(\phi)}& \gamma_k^{(\zeta)}& -\gamma_k^{(Z)} & \gamma_k^{(\chi)} \\
\gamma_k^{(\chi)}& -\gamma_k^{(Z)}& -\gamma_k^{(\zeta)} & -\gamma_k^{(\phi)} \\
\gamma_k^{(Z)}& \gamma_k^{(\chi)}& \gamma_k^{(\phi)} & -\gamma_k^{(\zeta)}
\end{pmatrix}.
\end{align}

As a crosscheck, it is instructive to inspect Eqs.\,(\ref{fullzz})--(\ref{fullzeta}) with respect to their frequency symmetries. Starting with $\chi_k$ and $\phi_k$, we know that the first summands of Eqs.\,(\ref{fullchi}) and (\ref{fullphi}), corresponding to the respective contributions due to the lowest order Feynman diagram, are even in $\omega_k$, since $\gamma_k^{(\chi)}$ and $\gamma_k^{(\phi)}$ are even in frequency. Further, in both equations the vertex corrections give rise only to products $\gamma_{k_1}\cdot\gamma_{k_2}\cdot\gamma_{k_3}$ that are even in $\omega_k$. This conclusion can be drawn by considering that $\gamma_k^{(\chi)}$, $\gamma_k^{(\phi)}$ are even, while $\gamma_k^{(Z)}$, $\gamma_k^{(\zeta)}$ are odd. The opposite holds when considering the functions $\omega_kZ_k$ and $\zeta_k$ in Eqs.\,(\ref{fullzz}) and (\ref{fullzeta}), respectively. Here, all first and second order contributions are odd along the Matsubara frequency axis, such that function $Z_k$ is even and $\zeta_k$ is odd.

We note further that we can connect the current equations to the theory of Section \ref{scEven} by setting $\zeta_k\equiv0$. As a consequence, also $\gamma_k^{(\zeta)}=0$ and we recover $\vec{\gamma}_k^T$ as 3-component vectors. Furthermore, for $f=Z,\chi,\phi$ we need to consider only the upper $3\times3$ submatrix of $Q_k^{(f)}$, set $\gamma_k^{(\zeta)}=0$ and obtain the matrix $P_k^{(f)}$ as introduced in Section \ref{scEven}. \\

In the following sections we apply the here-introduced theory to a cuprate model system. From here on, when referring to an induced state $\zeta_k^{\mathrm{ind}}$, we mean that the vertex-corrected Eliashberg equations of Section \ref{scEven} are solved self-consistently. Afterwards we calculate $\zeta_k^{\mathrm{ind}}$ from Eq.\,(\ref{zetainduced}) in a `one-shot' calculation. On the other hand, when referring to function $\zeta_k$, we mean the numerical solution from the four coupled Eliashberg equations in Section \ref{scCoexist}. All calculations are performed with the Uppsala Superconductivity Code (UppSC) \cite{UppSC,Aperis2015,Bekaert2018,Schrodi2020_3,Schrodi2020_4,Schrodi2021}, see also Appendix \ref{scAppComp} for numerical details.

\section{Cuprate model system}
\label{scCuprate}

We start the numerical analysis by considering a one-band tight-binding electron energy dispersion that is
 representative for the family of copper-based superconductors (cuprates). The electron energies are defined as
\begin{align}
\xi_{\mathbf{k}} = -t^{(1)} \big[\cos(k_x)+\cos(k_y)\big] - t^{(2)}\cos(k_x)\cos(k_y) - \mu , \label{xik}
\end{align}
with nearest and next-nearest neighbor hopping energies $t^{(1)}=0.25\,\mathrm{eV}$ and $t^{(2)}=-0.1\,\mathrm{eV}$. The chemical potential is fixed at $\mu=-0.09\,\mathrm{eV}$. This parameter choice renders a well-nested Fermi surface (compare Fig.\,\ref{comparezeta}) characteristic for the cuprates, see e.g.\ Refs.\,\cite{Tsuei2000,Damascelli2003} for a comprehensive overview. The Einstein phonon frequency and electron-phonon scattering strength are chosen here as $\Omega=50\,\mathrm{meV}$ and $g_0=150\,\mathrm{meV}$, respectively.

\subsection{Momentum dependence}\label{scMom}

\begin{figure}[h!]
	\centering
	\includegraphics[width=1\columnwidth]{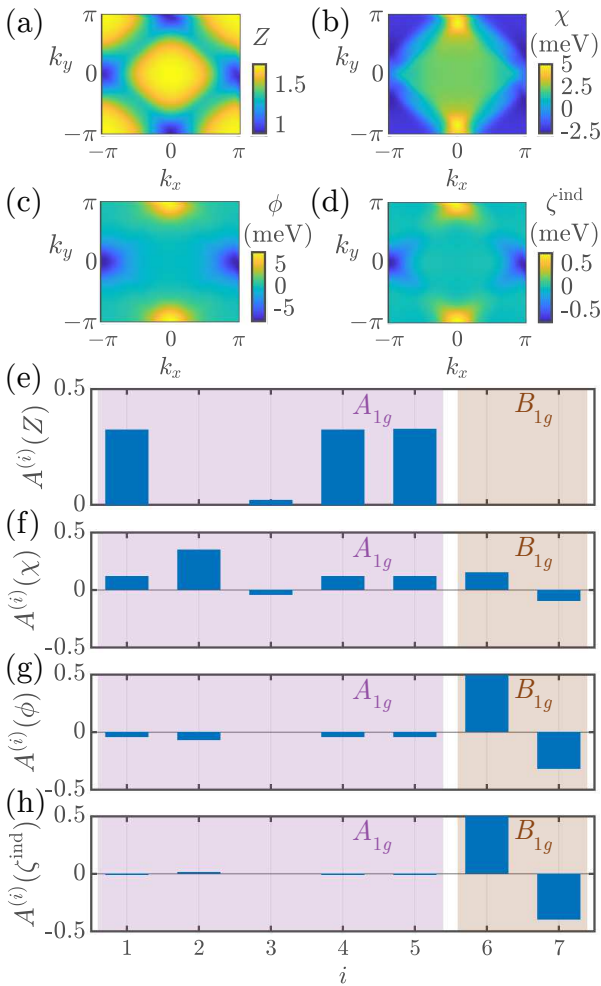}
	\caption{(a)-(c) Computed self-consistent solutions to Eqs.\,(\ref{z})--(\ref{phi}) for the mass renormalization $Z$ (a), chemical potential renormalization $\chi$ (b), and nematic even-frequency $d$-wave order parameter $\phi$ (c). (d) Induced odd-frequency $d$-wave spin-triplet order parameter. (e)-(h) Projections of the functions $Z$, $\chi$, $\phi$, and $\zeta^{\mathrm{ind}}$, according to form factors $f_{\mathbf{k}}^{(i)}$, with $i=1,\cdots \,,7$. $f_{\mathbf{k}}^{(1)}=1$, $f_{\mathbf{k}}^{(2)}=\cos(k_x)+\cos(k_y)$, $f_{\mathbf{k}}^{(3)}=\cos(k_x)\cos(k_y)$, $f_{\mathbf{k}}^{(4)}=\cos^2(k_x)+\cos^2(k_y)$, $f_{\mathbf{k}}^{(5)}=\sin^2(k_x)+\sin^2(k_y)$, $f_{\mathbf{k}}^{(6)}=\cos(k_x)-\cos(k_y)$, $f_{\mathbf{k}}^{(7)}=\cos(k_x)\cos(k_y)[\cos(k_x)-\cos(k_y)]$.}	\label{induced_t60}
\end{figure}

Here we analyze the momentum structure of zero-frequency components as found from Eqs.\,(\ref{z})--(\ref{phi}) and Eq.\,(\ref{zetainduced}) at $T=60\,\mathrm{K}$. In Fig.\,\ref{induced_t60}(a)--(c) we show our results for $Z=Z_{\mathbf{k},m=0}$, $\chi=\chi_{\mathbf{k},m=0}$ and $\phi=\phi_{\mathbf{k},m=0}$, respectively. As directly apparent from Fig.\,\ref{induced_t60}(b), the chemical potential renormalization has a nematic BZ structure, i.e., a mixture of $s$-wave and $s_{\pm}$-wave symmetry that breaks the $C_4$ to a $C_2$ rotational symmetry. The superconductivity order parameter $\phi$ has a $d$-wave shape with seemingly small nematicity \cite{SchrodiDWaveProj}.
% component.
 After solving Eqs.\,(\ref{z})--(\ref{phi}) self-consistently we calculate the induced odd-frequency spin-triplet order parameter via Eq.\,(\ref{zetainduced}). The result $\zeta^{\mathrm{ind}}=\zeta^{\mathrm{ind}}_{\mathbf{k},m=0}$ is shown in Fig.\,\ref{induced_t60}(d). Similarly as $\phi$ \cite{SchrodiDWaveProj}, $\zeta^{\mathrm{ind}}$ obeys $d$-wave symmetry arising from the nesting properties of $\xi_{\mathbf{k}}$. Additionally, we learn that the induced function $\zeta^{\mathrm{ind}}$ is approximately one order of magnitude smaller than $\phi$, and does not seem to have a nematic component. Note that the here-obtained ($d$-wave) odd-frequency order parameter is distinct from the odd-frequency $s$-wave, spin-triplet order parameter proposed originally by Berezinskii for $^3$He \cite{Berezinskii1974}.

For a closer investigation of the momentum symmetries we define seven different form factors, $f_{\mathbf{k}}^{(i)}\in\{1,~ \cos(k_x)+\cos(k_y),~ \cos(k_x)\cos(k_y),~ \cos^2(k_x)+\cos^2(k_y),~ \sin^2(k_x)+\sin^2(k_y),~ \cos(k_x)-\cos(k_y)$, $ \cos(k_x)\cos(k_y)[\cos(k_x)-\cos(k_y)] \}$, $i=1,\dots,7$. The form factors with indices $i=1,\dots,\,5$ belong to the $A_{1g}$ representation, while $i=6$ and $i=7$ belong to the $B_{1g}$ representation. Considering a function $h$, we can get the contribution in $h$ due to BZ dependence $f_{\mathbf{k}}^{(i)}$ by calculating the projection
\begin{align}
A^{(i)}(h) = \frac{\sum_{\mathbf{k}}f^{(i)}_{\mathbf{k}}h_{\mathbf{k},m=0}}{\sum_i\big|\sum_{\mathbf{k}}f^{(i)}_{\mathbf{k}}h_{\mathbf{k},m=0}\big|} .
\end{align}
In Fig.\,\ref{induced_t60}(e-h) we show the contributions due to each $f_{\mathbf{k}}^{(i)}$ for functions $Z$, $\chi$, $\phi$ and $\zeta^{\mathrm{ind}}$, respectively. Form factors belonging to the $A_{1g}$ ($B_{1g}$) representation are highlighted with purple (brown) background color. We have additionally tested other BZ symmetries, but the ones shown in Fig.\,\ref{induced_t60} are clearly the most relevant. 

From Fig.\,\ref{induced_t60}(e) we learn that the mass renormalization function has no finite projection in the $B_{1g}$ channel, which is why $Z$ is not nematic, compare also panel (a). Even though $\chi$ shows prevalent contributions from the $A_{1g}$ representation, the projections for $i=6$ and $i=7$ are non-negligible, which is why the chemical potential clearly breaks $C_4$ rotational symmetry. A similar picture emerges for the nematic order parameter $\phi$, see Fig.\,\ref{induced_t60}(g); here the dominant representation is $B_{1g}$, while small (but finite) contributions stem from $A_{1g}$. Turning to the induced term $\zeta^{\mathrm{ind}}$, the projections in the $A_{1g}$ channel are negligible to good approximation, which is why we observe a non-nematic $d$-wave state in Fig.\,\ref{induced_t60}(d).

\begin{figure}[th!]
	\centering
	\includegraphics[width=1\columnwidth]{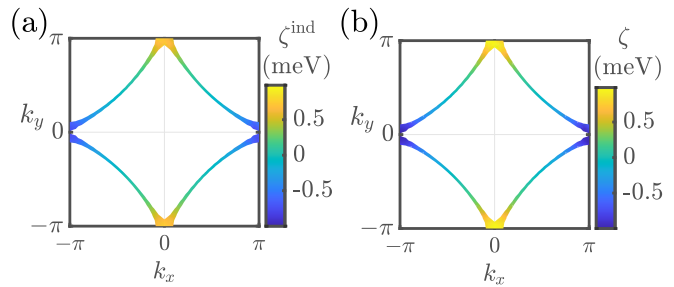}
	\caption{(a) Renormalized Fermi surface,  colored by $\zeta^{\mathrm{ind}}$ obtained from a single-shot calculation. (b) Same as (a), but using the fully self-consistent solutions to Eqs.\,(\ref{fullzz})--(\ref{fullzeta}).}	\label{comparezeta}
\end{figure}

So far we have shown that the induced order parameter is finite due to the vertex correction to Eliashberg theory, and results from a mixing of different momentum space symmetries. However, it remains to be shown that the actual (renormalized) interaction can support such an odd-frequency spin-triplet state, i.e., we need to solve the fully self-consistent Eqs.\,(\ref{fullzz})--(\ref{fullzeta}). We find that $\zeta_{k}$ is in fact very similar to the `one-shot' calculated $\zeta_k^{\mathrm{ind}}$. In Fig.\,\ref{comparezeta}(a) we show the induced function $\zeta^{\mathrm{ind}}$ as calculated before, projected on the renormalized Fermi surface, which is given by the condition
\begin{align}
\big[ \xi_{\mathbf{k}} + \chi_{\mathbf{k},m=0} \big] / Z_{\mathbf{k},m=0} = 0 . \label{fs}
\end{align}
On the other hand, using the fully self-consistent results for $\zeta=\zeta_{\mathbf{k},m=0}$ does hardly lead to significant changes, as can be seen in Fig.\,\ref{comparezeta}(b). Both functions obey $d$-wave symmetry, and are of similar magnitude. Our calculations also reveal that the remaining functions $Z$, $\chi$ and $\phi$ are hardly affected, depending on whether the odd-frequency spin-triplet order parameter is included self-consistently, or calculated in an isolated way.

\subsection{Temperature evolution}\label{scT}

In this section we discuss the temperature dependence of the superconducting state for our cuprate model system described by Eq.\,(\ref{xik}). For this purpose we define the experimentally observable superconducting gap $\Delta_{\mathbf{k}} = \phi_{\mathbf{k},m=0}/Z_{\mathbf{k},m=0}$. This function describes an energy gap due to even-frequency spin-singlet Cooper pairs. Analogously, we define the odd-frequency spin-triplet gap function as $\eta_{\mathbf{k}}=\zeta_{\mathbf{k},m=0}/Z_{\mathbf{k},m=0}$, or alternatively, $\eta_{\mathbf{k}}^{\mathrm{ind}}=\zeta_{\mathbf{k},m=0}^{\mathrm{ind}}/Z_{\mathbf{k},m=0}$ when we are concerned with the non-self-consistent induced order parameter $\zeta_{\mathbf{k},m=0}^{\mathrm{ind}}$. 

\begin{figure}[t!]
	\centering
	\includegraphics[width=1\columnwidth]{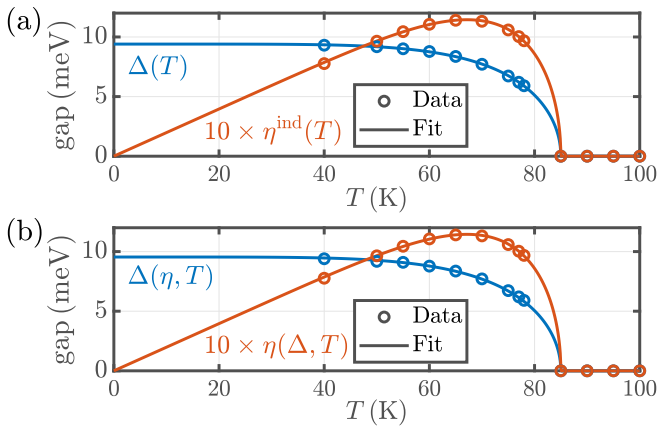}
	\caption{Temperature dependence of the even-frequency spin-singlet (blue) and the odd-frequency spin-triplet gap function (times ten) (red). Open circles represent our computed results, while solid lines are obtained from fitting our data to Eqs.\,(\ref{fitdelta}) and (\ref{fitzeta}). (a) Results obtained from Eqs.\,(\ref{z})--(\ref{phi}) and (\ref{zetainduced}). (b) Self-consistently computed temperature dependence from Eqs.\,(\ref{fullzz})--(\ref{fullzeta}).}	\label{tempdep}
\end{figure}

We start by numerically solving Eqs.\,(\ref{z})--(\ref{phi}) as function of temperature, which leads to the blue circles in Fig.\,\ref{tempdep}(a), representing $\Delta(T)=\underset{\mathbf{k}}{\mathrm{max}}|\Delta_{\mathbf{k}}(T)|$. For all temperatures at which $\Delta(T)$ is finite we find the same BZ symmetries as in Section \ref{scMom}. The functional behavior of the even-frequency superconducting gap is accurately modeled by
\begin{align}
\Delta(T) = \mathrm{Re} \sqrt{a - b\cdot  T^c} ,\label{fitdelta}
\end{align}
shown as solid blue curve. The variables $a$, $b$, and $c$ in Eq.\,(\ref{fitdelta}) are fitting parameters. The odd-frequency spin-triplet order parameter $\eta^{\mathrm{ind}}(T)=\underset{\mathbf{k}}{\mathrm{max}}|\eta^{\mathrm{ind}}_{\mathbf{k}}(T)|$, as calculated from Eq.\,(\ref{zetainduced}), is shown in red color in Fig.\,\ref{tempdep}(a). Again, open circles represent our numerical results, while the solid line is found from a functional fit, here given as
\begin{align}
\eta^{\mathrm{ind}}(T),\,\eta(T) = T \,\mathrm{Re} \sqrt{a - b\cdot  T^c} . \label{fitzeta}
\end{align}

As direct comparison to $\eta^{\mathrm{ind}}$ we also solved the fully self-consistent Eqs.\,(\ref{fullzz})--(\ref{fullzeta}) for $\eta$ as function of $T$, leading to the curves shown in Fig.\,\ref{tempdep}(b). 
It is directly apparent that results computed from the two different approaches only differ marginally. The zero-temperature limit of the even-frequency gap is in both cases fitted as $\Delta(T\rightarrow0)\simeq10\,\mathrm{meV}$, while the transition temperatures are given by $T_c\simeq85\,\mathrm{K}$. Further, we learn that the odd-frequency gap can equally accurately be determined in a single-shot or self-consistent calculation. This confirms the observations already made in Section \ref{scMom}, i.e., it is a good approximation to calculate $\zeta_k$ (or equivalently $\eta_k$) via Eq.\,(\ref{zetainduced}), requiring significantly less computational resources.

Our data reveals that the odd-frequency superconducting gap is approximately one order of magnitude smaller than its even-frequency counterpart. This finding is in agreement with previous adiabatic Eliashberg theory calculations of the odd-freuqency gap using \textit{ab initio} input \cite{Aperis2015,Aperis2020}. However, in contrast to these previous works where the odd-frequency order parameter gives rise to a paramagnetic Meissner effect, here the obtained odd-frequency superconductivity is expected to lead to a diamagnetic response. 
We suppose that this is the reason why it has been experimentally elusive so far, since both order parameters are expected to yield a diamagnetic Meissner effect. Therefore it is natural to associate the superconducting properties of the system with the dominant even-frequency order parameter only. Furthermore, it is apparent that $\eta\rightarrow0\,\mathrm{meV}$ for small temperatures, which is consistent with earlier predictions\,\cite{Balatsky1992,Fuseya2003}, while the even-frequency gap approaches a finite value $\Delta(0)$ as $T\rightarrow0\,\mathrm{K}$. Note, that Fig.\,\ref{tempdep} contains few points for small temperatures, which stems from the necessary increase in the number of Matsubara frequencies to ensure convergence, becoming computationally unfeasible in the limit $T\rightarrow0\,\mathrm{K}$, see also Appendix \ref{scAppComp}.

\section{Summary and Discussion}\label{scSummary}

We have shown that vertex corrections to the electron-phonon problem in superconductors generally can lead to a coexistence of even- and odd-frequency superconducting states. Our analysis reveals that, due to a mixing of momentum space representations, the dominant spin-singlet even-frequency order parameter induces a spin-triplet odd-frequency superconducting gap, which, in comparison, falls short approximately one order of magnitude in amplitude. Both order parameters have the same critical temperature and are expected to yield a diamagnetic Meissner effect, hence, there is no straightforward way of verifying the induced odd-frequency spin-triplet Cooper pairs experimentally. 

Our numerical results confirm the existence of a finite odd-frequency $d$-wave order parameter in our cuprate model system, serving as a proof of concept. We cannot rule out that a similar picture can be found in other classes of superconductors, hence,  it is possible that the induced state generically exists in a large variety of materials. Even though it has been argued in the past that odd-frequency superconductivity might only be realized for 
very anisotropic electron-phonon interaction mediated by acoustic phonons \cite{Balatsky1992}, or, alternatively, for spin-dependent couplings \cite{Abrahams1993}, our results are obtained for a completely isotropic electron-phonon coupling mediated by optical phonons. Our results are enabled by the renormalized vertex function, see Refs.\,\cite{Schrodi2020_2,SchrodiDWaveProj,SchrodiThFeAsN}, which is closely associated with Fermi surface nesting conditions. This allows for a self-consistent and strongly anisotropic interaction, supporting both even- and odd-frequency superconducting gaps to coexist.
Our results furthermore imply that a stronger realization of an odd-frequency order parameter can be expected in nonadiabatic superconductors where vertex corrections beyond Migdal's approximation are important.

Lastly, it is currently an open question whether systems with \textit{dominant} odd-frequency order parameter exist. We speculate that, within the formalism used here, this is an unlikely scenario because the primary even-frequency gap induces its odd-frequency counterpart. However, if we would assume odd-momentum space parity, the situation would be different. In this case, $\phi$ would still be dominant and describe spin-singlet odd-frequency Cooper pairs with e.g.\ $p$-wave BZ symmetry, while $\zeta$ would represent an induced spin-triplet even-frequency order parameter, a scenario that we leave for future investigations.

\begin{acknowledgments}
This work has been supported by the Swedish Research Council (VR), the R{\"o}ntgen-{\AA}ngstr{\"o}m Cluster, and the Knut and Alice Wallenberg Foundation (grant No.\ 2015.0060). 
The Eliashberg-theory calculations were enabled by resources provided by the Swedish National Infrastructure for Computing (SNIC) at NSC Link\"oping, partially funded by VR through Grant Agreement No.\ 2018-05973.
\end{acknowledgments}

\appendix

\section{Computational details}\label{scAppComp}

All numerical results presented in this work have been carried out with at least 32 momentum points along each spatial dimension. We performed convergence checks with larger grids to ensure that our calculations are reliable. The number of Matsubara frequencies $\omega_m$, $q_l$, with $m,l\in[-\mathcal{M},\mathcal{M}-1]$ was chosen as function of temperature, but always $2\mathcal{M}\gtrsim250$. The vertex-corrected Eliashberg equations are solved by using Fourier convolution techniques, which is a crucial aspect of our implementation for the simulations to be computationally feasible. For this purpose all self-consistent functions must be interpolated to a larger frequency grid $[-3\mathcal{M},3\mathcal{M}-1]$ at the beginning of each step of the self-consistency cycle, see also Ref.\,\cite{Schrodi2020_2}. As convergence criterion, we set the threshold to the maximally allowed absolute change in $Z$, $\chi$, $\phi$ (and $\zeta$, if included self-consistently) to $10^{-7}$.

Usually, when taking the odd-frequency spin-triplet order parameter $\zeta_k$ self-consistently into account, i.e.\ solving Eqs.\,(\ref{fullzz})--(\ref{fullzeta}) instead of Eqs.\,(\ref{z})-(\ref{phi}) and (\ref{zetainduced}), the number of iterations needed for convergence increases significantly. It has been shown in the main text that it is often a good approximation to use $\zeta^{\mathrm{ind}}_k$ instead of $\zeta_k$ to reduce the run time, as both BZ symmetry and order of magnitude do not change.

%\newpage

\bibliographystyle{apsrev4-1}


%\bibliography{bib.bib}{}

\end{document}